\documentclass[12pt,superscriptaddress]{revtex4}

\usepackage{amsmath}
\usepackage{graphicx}
\usepackage{subfigure}

\begin{document}

\title{Kinetic theory of electromagnetic ion waves in relativistic plasmas}

\author{Mattias Marklund}
\altaffiliation[Also at: ]{Centre for Fundamental Physics, Rutherford Appleton Laboratory,
  Chilton, Didcot, Oxon OX11 OQX, U.K.}
\affiliation{Centre for Nonlinear Physics, Department of Physics, Ume{\aa} University, 
SE--901 87 Ume{\aa}, Sweden}

\author{Padma K. Shukla}
\altaffiliation[Also at: ]{Centre for Fundamental Physics, Rutherford Appleton Laboratory,
  Chilton, Didcot, Oxon OX11 OQX, U.K.}
\altaffiliation{Centre for Nonlinear Physics, Department of Physics, Ume{\aa} University, 
SE--901 87 Ume{\aa}, Sweden}
\altaffiliation{Department of Physics, University of Strathclyde, Glasgow, Scotland, 
  G4 ONG, UK}
\altaffiliation{Centro de F\'{i}sica dos Plasmas, Instituto Superior T\'{e}cnico, 1049-001 Lisboa, 
  Portugal}
\affiliation{Institut f\"ur Theoretische Physik IV and Centre for Plasma Science 
  and Astrophysics, Fakult\"at f\"ur Physik und Astronomie, Ruhr-Universit\"at Bochum, 
  D--44780 Bochum, Germany}

\date{\today}

\begin{abstract}
  A kinetic theory for electromagnetic ion waves in a cold relativistic plasma is derived. The kinetic 
  equation for the broadband electromagnetic ion waves is coupled to the slow density response via 
  an acoustic equation driven by ponderomotive force like term linear in the electromagnetic field amplitude. 
  The modulational instability growth rate is derived for an arbitrary spectrum of waves. The monochromatic 
  and random phase cases are studied.  
\end{abstract}

\maketitle

The interaction of intense radiation with plasmas is of fundamental importance in a wide variety of applications, 
such as inertial confinement fusion \cite{mourou-etal,marklund} and pulsar emissions \cite{beskin-etal}, and such 
interactions can lead to a range of instabilities, e.g.\ Brillouin and Raman scattering as well as 
modulational instabilities \cite{yu,shukla,sharma,shukla-etal}. In some of the instabilities, relativistic 
effects play an important role \cite{shukla-etal,tsintsadze-stenflo,max-etal}.  In Refs.\ \cite{kotsarenko,st} 
it has been shown that a novel type of ion electromagnetic wave may propagate through a cold magnetized 
relativistic plasma. Such waves may be of relevance in high-energy astrophysical environments, e.g.\ pulsars, 
as well as in intense laser interactions with high density targets. However, the coherence length of intense 
electromagnetic (EM) waves could often be short. Moreover, it is well-known that effects of partial coherence 
may be used to stabilize the propagation of EM pulses in nonlinear dispersive media \cite{kato-etal,koenig-etal}. 
Thus, it is of interest to analyze the effects of partial coherence of EM ion wave propagation. 

In this paper, the effects of partial coherence ion EM waves in a magnetized relativistic plasma is analyzed. 
In particular, a kinetic equation describing the evolution of the EM quasi-particles is derived and the general 
nonlinear dispersion relation is obtained. We study the effects of partial coherence using a random phase 
background of EM ion waves, leading  to a Lorentz distribution for the quasi-particles. It is shown that the 
spatial spectral broadening induced by the random phase gives rise to a reduction of the modulational instability 
growth rate, in addition to the appearance of a new short wavelength instability region. 

Let us consider large amplitude circularly polarized electromagnetic ion waves, with the electric field 
amplitude $E_0$, propagating parallel to a constant magnetic field $\vec{B}_0 = B_0\hat{z}$ in a cold 
electron--ion plasma.  Stenflo and Tsintsadze \cite{st} have shown that there are situations in which 
the ions are strongly relativistic. For such electromagnetic waves, the linear dispersion relation 
is of the form

\begin{equation}\label{eq:lindisp}
  1 \approx \frac{k^2c^2}{\omega^2} \pm \frac{\omega_{\mathrm{p}i}^2}{\omega\omega_{Ei}} ,
\end{equation}
where $k$ is the wavenumber, $\omega$ is the wave frequency, $\omega_{\mathrm{p}i} 
= (e^2n_{0i}/\epsilon_0m_i)^{1/2}$ is the ion plasma frequency, $\omega_{Ei} = eE_0/cm_i$, $e$ is 
the magnitude of the electron charge, $n_{0i}$ is the unperturbed ion density, $\epsilon_0$ is the 
vacuum dielectric constant, $m_i$ is the ion rest mass, and $c$ is the speed of light in vacuum. 
The $+$ ($-$) in (\ref{eq:lindisp}) correspond to $|\vec{p}_i|/m_ic > \omega_{\mathrm{c}i}/\omega$ 
($|\vec{p}_i|/m_ic < \omega_{\mathrm{c}i}/\omega$), where $\vec{p}_i$ is the ion particle momentum 
and $\omega_{\mathrm{c}i} = eB_0/m_i$ is the ion cyclotron frequency. If the plus sign is used in 
the dispersion relation (\ref{eq:lindisp}), we obtain the so called second ion-cyclotron wave, 
whereas the minus sign gives the ion helicon wave.  

The nonlinear evolution of electromagnetic ion waves, with amplitude $E$, in a cold quasi-neutral 
relativistic plasma is given by \cite{st}

\begin{equation}\label{eq:nlse}
  i\left(\frac{\partial}{\partial t} + v_g\frac{\partial}{\partial z}\right)E 
  + \frac{v_g'}{2}\frac{\partial^2 E}{\partial z^2} 
    \pm \omega\left( \frac{|E| - E_0}{E_0} - \frac{\delta n}{n_0} \right)E = 0 ,  
\end{equation}
where the $+$ ($-$) refers to the $+$ ($-$) in the linear dispersion relation (\ref{eq:lindisp}), 
and the slow density response is obtained from 

\begin{equation}\label{eq:acoustic}
  \left( \frac{\partial^2}{\partial t^2} - v_s^2\frac{\partial^2}{\partial z^2}\right)\frac{\delta n}{n_0} 
    = \pm\beta\frac{\partial^2}{\partial z^2}\left(\frac{|E| - E_0}{E_0}\right) .
\end{equation}
Here $v_g = \partial\omega/\partial k$ is the group velocity, $v_g' = \partial^2\omega/\partial k^2$ is the 
group velocity dispersion, $\delta n$ is the ion density perturbation, $n_0$ is the constant background density, 
$v_s = (T_e/m_i)^{1/2}$ is the ion sound speed, and $\beta = (c^2/2)(E_0/cB_0)$. Thus, the above system 
of equations is effectively a quadratic nonlinear Schr\"odinger equation, in contrast to the regular cubic 
nonlinear Schr\"odinger equation. 

In the quasi-stationary limit, we may integrate Eq.\ (\ref{eq:acoustic}) to obtain the quadratic nonlinear 
Schr\"odinger equation

\begin{equation}\label{eq:qs}
   i\frac{\partial E}{\partial\tau} + \frac{v_g'}{2}\frac{\partial^2 E}{\partial\zeta^2} 
    \pm \omega\left( 1 \pm \frac{\beta}{v_s} \right)\frac{|E|}{E_0}E = 0 ,
\end{equation}
where we have transformed to a co-moving frame $\tau = t$ and $\zeta = z - v_gt$.

With the ansatz $E = (E_0 + \delta E)\exp(i\phi)$, where $\delta E \ll E_0$ and $\phi \ll 1$, we linearize 
Eqs.\ (\ref{eq:nlse}) and (\ref{eq:acoustic}). For a harmonic dependence, i.e.\ $\delta E, \phi \propto 
\exp(iKz - i\Omega t)$, we obtain the dispersion relation

\begin{equation}\label{eq:mono-disp}
  (\Omega^2 - v_s^2K^2)\left[ (\Omega - v_gK)^2 \pm \tfrac{1}{2}\omega v_g'K^2 - \tfrac{1}{4}v_g'^2K^4\right]
    = \tfrac{1}{2}\omega\beta v_g'K^4 ,
\end{equation}
consistent with the results in Refs.\ \cite{st} and \cite{tsintsadze-etal}. In the quasi-stationary limit 
(cf.\ Eq.\ (\ref{eq:qs})), we can solve for $\Omega$ to obtain the growth rate $\Gamma = \mathrm{Im}(\Omega)$ 
according to

\begin{equation}
  \Gamma  = K\left[ \frac{1}{2}\omega v_g' \left(\frac{\beta}{v_s^2}  \pm 1\right) - \frac{1}{4}v_g'^2K^2
    \right]^{1/2} .
\end{equation}
For positive group velocity dispersion, we see that the second order dispersive term competes with the nonlinear 
term, giving the characteristic modulational instability growth rate curve (cf.\ Figs.\ 1 and 2).  

We are now interested in analyzing the effects of partial coherence of the electromagnetic ion waves. 
For this purpose, we introduce the Wigner function \cite{wigner,moyal,klimontovich,mendonca}

\begin{equation}\label{eq:wignerfunc}
  \rho(t,z,p) = \frac{1}{2\pi}\int d\zeta\,e^{ip\zeta}E^*(t,z + \zeta/2)E(t,z - \zeta/2),
\end{equation}
such that 

\begin{equation}\label{eq:norm}
  |E| = \left( \int dp\,\rho(t,z,p)\right)^{1/2} .
\end{equation}
We note that the Wigner method \cite{anderson-etal}, as well as 
the equivalent mutual coherence method \cite{demetrios-etal},
have been used to analyze the modulational instability of the
cubic nonlinear Schr\"odinger equation relevant for e.g.\ nonlinear optics.
Moreover, in plasma applications the Zakharov equations have been analyzed 
using the above method \cite{renato-etal} in order to obtain the statistical dynamics and Landau like damping of
Langmuir waves. 

Applying the time derivative to the definition (\ref{eq:wignerfunc}) and using (\ref{eq:nlse}), 
we obtain the kinetic equation

\begin{equation}\label{eq:kinetic}
  \frac{\partial\rho}{\partial t} + \left( v_g + v_g'p \right)\frac{\partial\rho}{\partial z} 
    \pm 2\omega\left( \frac{|E|}{E_0} - \frac{\delta n}{n_0} \right)
    \sin\left( \frac{1}{2}\stackrel{\leftarrow}{\frac{\partial}{\partial z}}
    \stackrel{\rightarrow}{\frac{\partial}{\partial p}}\right)\rho = 0 ,
\end{equation}
where the $\sin$-operator is determined in terms of its Taylor expansion and the arrows denotes 
direction of operation.  The kinetic equation (\ref{eq:kinetic}), together with (\ref{eq:acoustic}) 
and (\ref{eq:norm}), determines the evolution of broad band electromagnetic ion waves in cold relativistic plasmas. 

In order to analyze the modulational instability of the system (\ref{eq:acoustic}), (\ref{eq:norm}), 
and (\ref{eq:kinetic}), we let $\rho(t,z,p) = \rho_0(p) + \delta\rho\,\exp(iKz - i\Omega t)$ and 
$\delta n \propto \exp(iKz - i\Omega t)$, where $\delta\rho \ll \rho_0$. Linearizing 
Eqs.\ (\ref{eq:acoustic}), (\ref{eq:norm}), and (\ref{eq:kinetic}), we then obtain the nonlinear 
dispersion relation

\begin{equation}\label{eq:disprel}
  E_0^2 = \pm \frac{\omega}{2}\left( 1 \mp \beta\frac{K^2}{\Omega^2 - v_s^2K^2} \right)
    \int dp\,\frac{\rho_0(p + K/2) - \rho_0(p - K/2)}{\Omega - (v_g + v_g'p)K} ,
\end{equation}
where $E_0 = (\int dp\,\rho_0)^{1/2}$.
This dispersion relation generalizes (\ref{eq:mono-disp}) to the case of arbitrary spatial spectral 
background distributions $\rho_0$. 

In the case of a monochromatic spectral distribution, $\rho_0(p) = E_0^2\delta(p)$, we retrieve the 
dispersion relation (\ref{eq:mono-disp}), as expected. However, if the background distribution $\rho_0$ 
has a finite spectral width, the dispersion relation is altered. Next, we look at the case of a random phased 
background electromagnetic ion wave. This will give rise to the spectral distribution in the form of 
the Lorentzian

\begin{equation}
  \rho_0(p) = \frac{E_0^2}{\pi}\frac{\Delta}{p^2 + \Delta^2} ,
\end{equation}
where $\Delta$ is the spectral width of the distribution. With this, the nonlinear dispersion relation 
(\ref{eq:disprel}) becomes

\begin{equation}\label{eq:disp-lorentz}
  (\Omega^2 - v_s^2K^2)\left[ [\Omega - (v_g - iv_g'\Delta)K]^2 \pm \tfrac{1}{2}\omega v_g'K^2 
  - \tfrac{1}{4}v_g'^2K^4\right]
    = \tfrac{1}{2}\omega\beta v_g'K^4 .
\end{equation}
As $\Delta \rightarrow 0$, we obtain from above the monochromatic dispersion relation (\ref{eq:mono-disp}). 
Moreover, the effect of the spectral broadening is to introduce a damping of the perturbation modes. 

Next, we analyze the modulational instability properties of the dispersion relation (\ref{eq:disp-lorentz}). 
We introduce the dimensionless variables $\bar{\Omega} = \Omega/\omega$, $\bar{K} = v_gK/\omega$, $\bar{\Delta} 
= v_g\Delta/\omega$, $\bar{v} = v_s/v_g$, $\bar{v}' = \omega v_g'/v_g^2$, and $\bar{\beta} = \beta/v_g^2$. 
In Figs.\ 1 ($+$ in Eq.\ (\ref{eq:nlse})) and 2 ($-$ in Eq.\ (\ref{eq:nlse})) we have displayed the growth 
rate $\Gamma = \mathrm{Im}({\bar{\Omega}})$ for some typical parameter values. Using $\bar{v} = 0.33$, 
$\bar{\beta} = 1$, and $\bar{v}' = 0.5$, it can be seen that a finite spectral width $\bar{\Delta}$ gives 
rise to a reduced growth rate. On the other hand, the $\bar{K}$-region, where the instability occurs, is enlarged. 
We also note that the $+$-mode has a larger growth rate as compared to the $-$-mode in Eq.\ (\ref{eq:kinetic}).

\begin{figure}
  \subfigure[]{\includegraphics[width=0.48\textwidth]{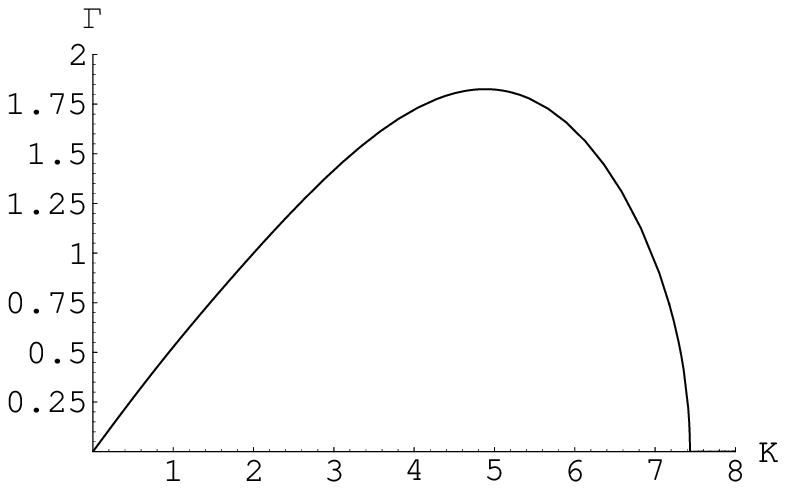}}
  \subfigure[]{\includegraphics[width=0.48\textwidth]{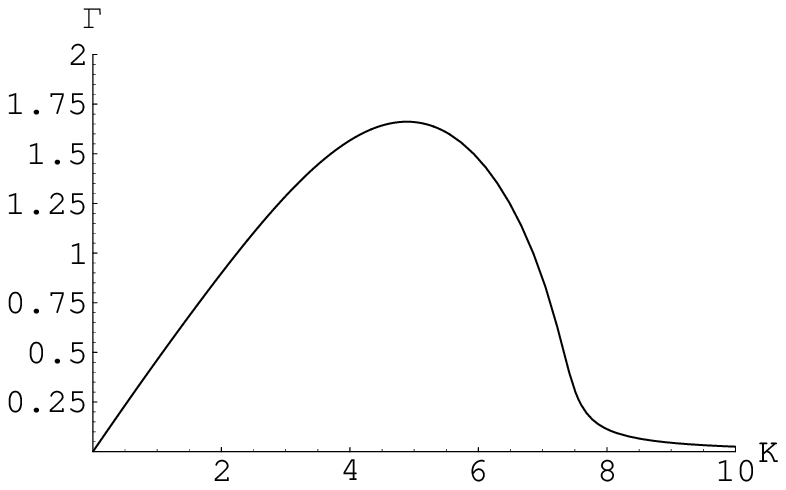}}
  \caption{The growth rate $\Gamma = \mathrm{Im}(\bar{\Omega})$ plotted as a function of the normalized 
  wavenumber $\bar{K}$ for the parameter values $\bar{v} = 0.33$, $\bar{\beta} = 1$, and $\bar{v}' = 0.5$, 
  with $+$ in Eq.\ (\ref{eq:nlse}). In (a) we have $\bar{\Delta} = 0$, while (b) has $\bar{\Delta} = 0.2$. 
  The reduction in growth rate can be seen from (a) to (b).}
\end{figure}

\begin{figure}
  \subfigure[]{\includegraphics[width=0.48\textwidth]{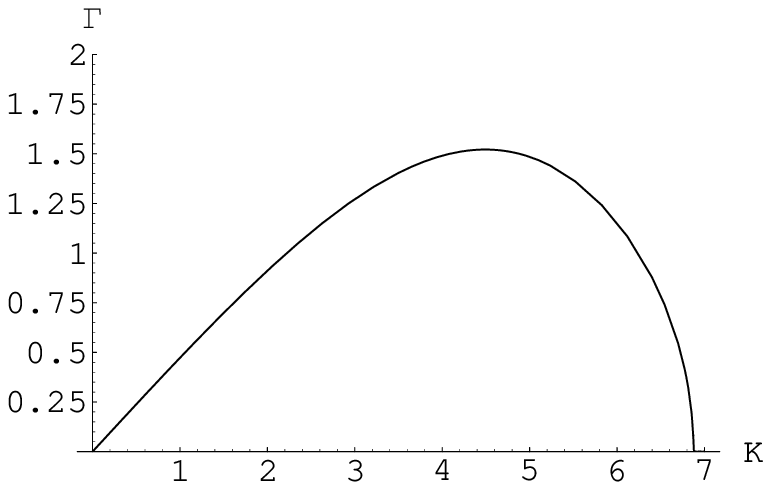}}
  \subfigure[]{\includegraphics[width=0.48\textwidth]{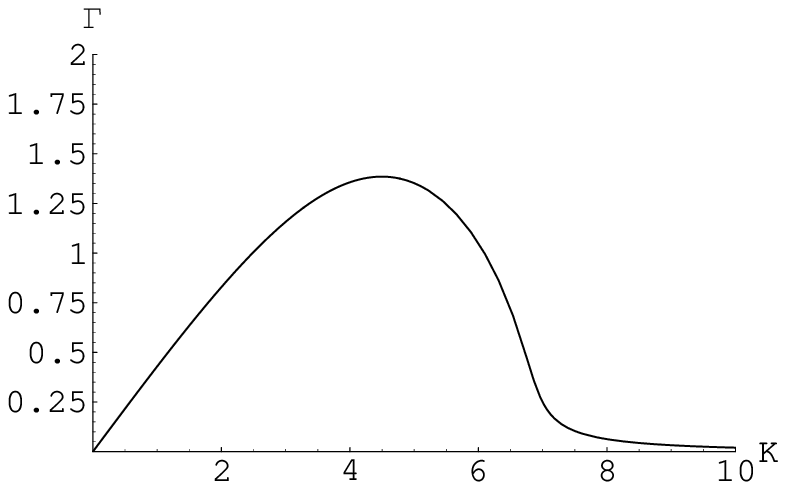}}
  \caption{The growth rate $\Gamma = \mathrm{Im}(\bar{\Omega})$ plotted as a function of the normalized 
  wavenumber $\bar{K}$ for the parameter values $\bar{v} = 0.33$, $\bar{\beta} = 1$, and $\bar{v}' = 0.5$, 
  with $-$ in Eq.\ (\ref{eq:nlse}). In (a) we have $\bar{\Delta} = 0$, while (b) has $\bar{\Delta} = 0.2$. 
  The reduction in growth rate can be seen from (a) to (b).}
\end{figure}

To summarize, we have analyzed the effects of a partial coherence of circularly polarized electromagnetic 
ion waves in relativistic plasmas. In particular, the modulational instability growth rate was found for both 
coherent waves and for waves with a random phase. It was shown that the effect of partial coherence is to 
stabilize the propagation of the aforementioned waves, while broadening the possible instability wavenumber region. 
The latter could lead to the formation of short wavelength nonlinear structures due to partial coherence. 

\acknowledgments
This research was partially supported by the Swedish Research Council. We thank L. Stenflo for helpful 
discussions and valuable insights.

\end{document}